\documentclass[conference]{IEEEtran}

\usepackage{cite}
\usepackage{amsmath,amssymb,amsfonts}
\usepackage{graphicx}
\usepackage{textcomp}
\usepackage{xcolor}

\usepackage{booktabs}
\usepackage{multirow}
\usepackage{siunitx}
\usepackage{tabularx}
\usepackage[table]{xcolor}
\usepackage{colortbl}

\usepackage{tikz}
\usetikzlibrary{shadows,arrows.meta,positioning}

\usepackage{hyperref}

\begin{document}

\title{Toward Real-Time Circadian Phase Estimation with Low Latency from Wearable Sensing Data}

\author{
\IEEEauthorblockN{Mengzhu Xu\IEEEauthorrefmark{1},
Nemanja Cabrilo\IEEEauthorrefmark{1},
Merel van Gilst\IEEEauthorrefmark{1}\IEEEauthorrefmark{2},
Jean-Paul Linnartz\IEEEauthorrefmark{1}}
\IEEEauthorblockA{\IEEEauthorrefmark{1}Eindhoven University of Technology, Eindhoven, The Netherlands}
\IEEEauthorblockA{\IEEEauthorrefmark{2}Center for Sleep Medicine Kempenhaeghe, Heeze, The Netherlands}
}

\maketitle
\begin{abstract}
Accurate estimation of the human circadian phase plays an important role in personalized health monitoring, but most existing wearable-based approaches operate retrospectively and require full circadian cycle recordings, leading to high estimation latency and substantial data and computational burden for real-time deployment on edge devices.

In this study, we investigated whether circadian phase can be estimated in real time using only short historical windows of wearable data. We propose a low latency framework that estimates instantaneous circadian phase from past observations, with a cosinor-fitted core body temperature rhythm serving as the reference.

Data from a free-living field study involving 14 participants were used to systematically evaluate the effects of sensor modality selection, historical window length, and model class under participant-based cross-validation. The results showed that estimation accuracy improves with increasing window length but saturates at approximately 8 hours of history. Tree-based models reached a performance plateau beyond 480 minutes, whereas sequence-based models continued to benefit from longer temporal contexts. When relying solely on light exposure and physical activity, the proposed approach achieved a mean circular mean absolute error (CMAE) of 1.19 h.

These findings provide practical guidance for efficient and deployable real-time circadian phase monitoring using wearables.
\end{abstract}


\section{Introduction}
The circadian rhythm is an intrinsic biological oscillation with a near 24-hour period, regulated by the suprachiasmatic nucleus and governing essential physiological and behavioral processes, including sleep-wake regulation, thermoregulation, and hormonal secretion~\cite{Zee2013}. Although self-sustained, it is continuously entrained by environmental cues, with light acting as the primary Zeitgeber that aligns internal biological time with the solar day~\cite{Czeisler1989}. Modern lifestyles increasingly disrupt this alignment due to insufficient daytime light exposure and excessive  evening artificial light~\cite{Klepeis2001}. Together with social constraints such as fixed work or school schedules, these factors lead to a mismatch between biological and social time, referred to as social jetlag~\cite{Wittmann2006}, which has been associated with adverse health outcomes, including metabolic, cardiovascular, cognitive, and immune dysfunction~\cite{Brainard2015,Baron2014}.

Dim light melatonin onset (DLMO) is a common reference for circadian phase~\cite{PandiPerumal2007}, but is impractical for real-world, real-time use due to controlled lighting, repeated sampling, and offline analysis requirements. Core body temperature (CBT) provides a less invasive alternative, with its nadir serving as a validated circadian phase marker~\cite{Klerman2002,Reid2019}. Advances in ingestible and wearable sensors now enable continuous CBT monitoring outside laboratory settings~\cite{Byrne2007,Marchiano2025}.

Recent advances in wearable sensing enable continuous collection of multimodal physiological and environmental data (e.g., light exposure, physical activity, skin temperature, and cardiovascular signals). Prior work has leveraged these signals to infer circadian phase in free-living settings using mathematical and machine-learning methods~\cite{Hannay2020,Bonarius2021ParticleFilterCircadian,Suarez2021CircadianPhasePrediction}. Across different sensing modalities and reference markers, these studies consistently report accuracies in the range of one to two hours~\cite{Stone2019,Sim2017}. Similar performance has been reported with alternative inputs such as metabolomics~\cite{Woelders2023Machine}. Errors of 1--2 h have been widely considered sufficient for a variety of downstream applications, including healthy buildings~\cite{Papatsimpa2020a,Hosseini2024Architectural}, chronotherapy~\cite{Levi2007}, personalized sleep~\cite{VanDongen2003,Papatsimpa2020b}, and performance optimization~\cite{Teo2011,Vitale2017}, where interventions operate on multi-hour timescales.

Despite these advances, most existing methods operate retrospectively, requiring data spanning one or more complete circadian cycles to identify a phase marker (e.g., DLMO or CBT nadir) and reconstruct the full rhythmic trajectory~\cite{Brown2021,Weed2025ParticleFiltering}. Although effective for offline analysis~\cite{Stone2020,huang2021predicting}, this strategy inherently introduces estimation latency, as phase marker identification requires observations extending beyond its occurrence. Phase estimates at the current time thus depend on future data, posing a fundamental challenge for real-time tracking and limiting applicability in resource-constrained wearable systems.

The contributions of this paper include:
\begin{itemize}
\item We investigate the feasibility of low latency circadian phase estimation from wearable data under free-living conditions.
\item We propose a framework that estimates instantaneous circadian phase using only historical observations, with a cosinor-fitted CBT rhythm as reference.
\item We systematically evaluate the effects of wearable modality combinations and window length on performance.
\item We analyze trade-offs between window length, model complexity, and estimation accuracy across different modeling approaches.
\end{itemize}
Overall, this work identifies practical and computationally efficient strategies for real-world circadian phase monitoring and supports future applications in adaptive lighting, chronotherapy, precision medicine, and performance optimization.

\section{Material and Data Collection}

\subsection{Subjects}
The data were collected from 14 healthy participants during a field study conducted between 2021 and 2022 under free-living conditions in the Netherlands. All recordings were performed within a single stable daylight season for each participant and did not overlap the transition between winter and summer time.

\begin{figure}[!ht]
\centering
\includegraphics[width=0.9\linewidth]{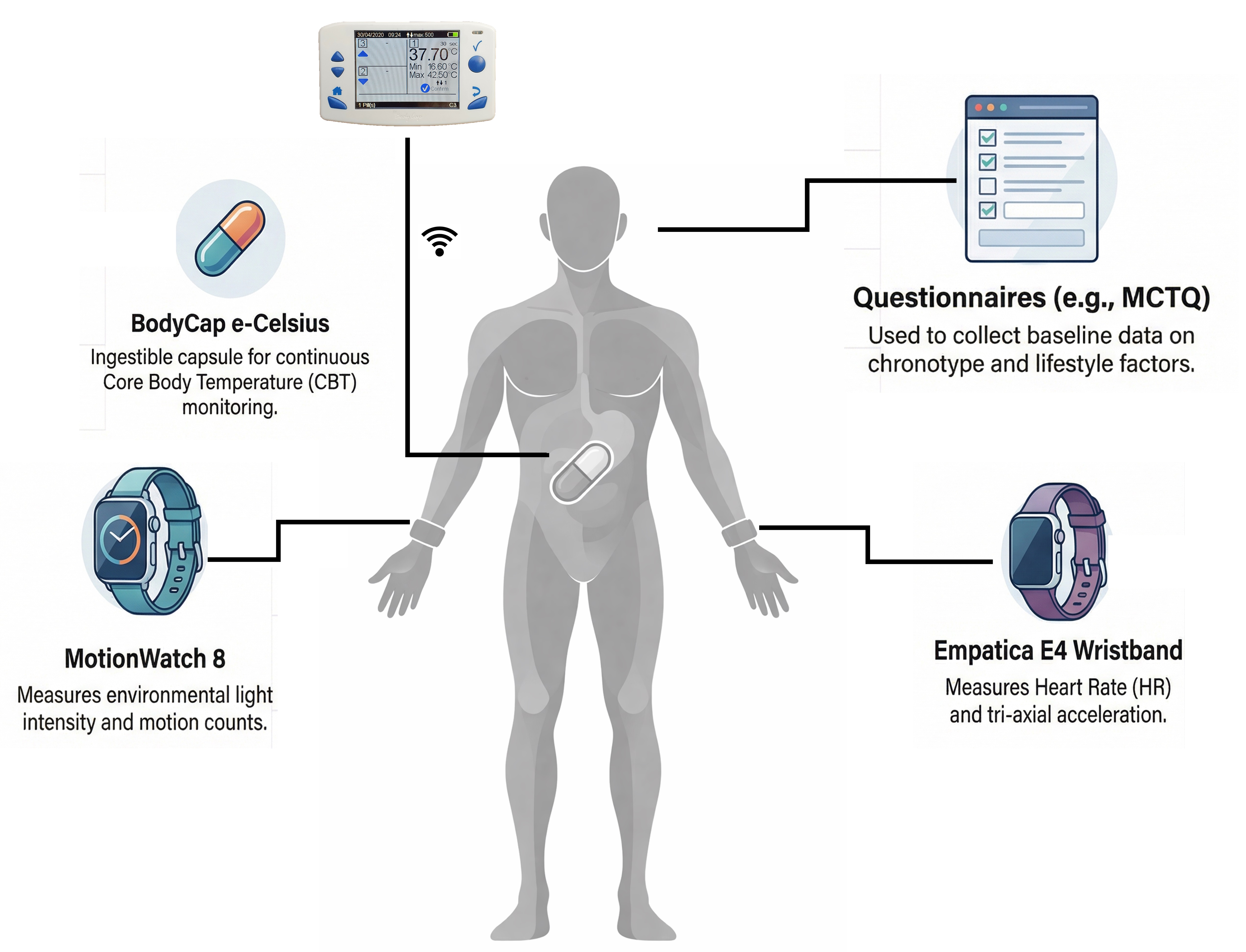}
\caption{Overview of the sensing setup used in the study.}
\label{fig:sensor_overview}
\end{figure}

Participants were continuously monitored using multiple wearable and ingestible sensors (Fig.~\ref{fig:sensor_overview}), allowing natural daily activity while providing multimodal information for circadian phase estimation. To reduce potential confounders that may influence circadian rhythms, individuals were excluded if they had diagnosed medical or sleep disorders, consumed more than three units of alcohol per week, more than three cups of coffee per day, more than three cigarettes per day, or worked irregular or shift schedules. All included individuals were working adults between 18 and 50 years of age, and no pregnant participants were enrolled.

All participants provided written informed consent. Chronotype and sleep-related questionnaires were used for screening only and were not included as model inputs.

\subsection{Data Collection and Schedule}
During the 20-day field protocol, all participants continuously wore devices measuring actigraphy, light exposure, skin temperature and heart rate (HR). CBT was scheduled to be measured during the last three days using two ingestible sensors in sequence to cover the final three days.

The main sensing modalities are summarized below.

\begin{itemize}
    \item \textbf{MotionWatch 8 (actigraphy and light):}  
    Worn on the wrist, the device measured environmental light intensity (lux) and wrist activity counts at a 1-minute resolution.
    
    \item \textbf{Skin temperature logger (DSL1922L):}  
    Placed on the wrist adjacent to the MotionWatch, the logger measured distal skin temperature every 5 minutes.
    
    \item \textbf{Empatica E4 wristband:}  
    Worn on the non-dominant wrist, the device recorded HR and tri-axial acceleration signals, which were aggregated to a 1-minute resolution.
    
    \item \textbf{Ingestible CBT sensor (BodyCap e-Celsius):}  
    Participants ingested electronic capsules that wirelessly transmitted measurements to a receiver at 5-minute intervals.
\end{itemize}

\section{Methods}

\subsection{Overview}

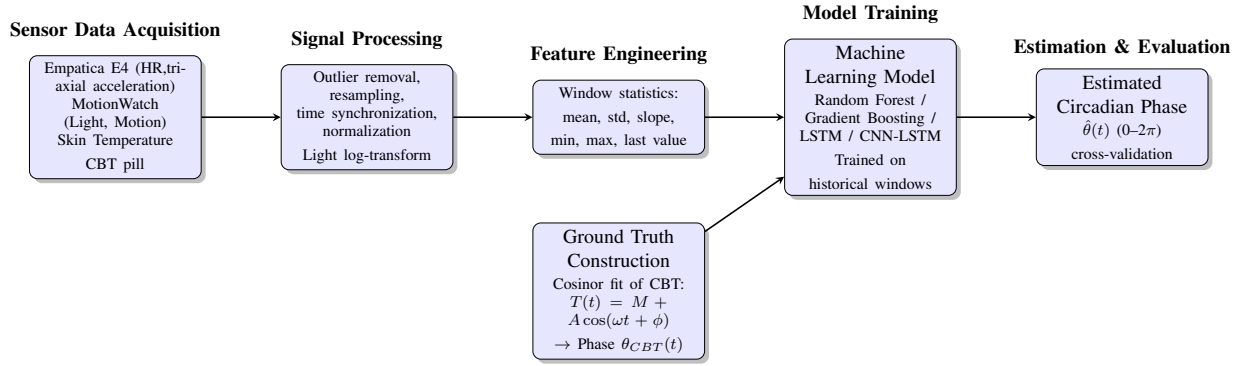
\begin{figure*}[t]
\centering
\resizebox{0.9\linewidth}{!}{%
\begin{tikzpicture}[node distance=1.2cm, every node/.style={align=center}]
\tikzstyle{process} = [rectangle, rounded corners, draw=black, fill=blue!10, drop shadow, text width=2.8cm, minimum height=1.0cm]
\tikzstyle{arrow} = [thick,->,>=stealth]

\node[process] (input) {
{\footnotesize Empatica E4 (HR,tri-axial acceleration)\\MotionWatch (Light, Motion)\\Skin Temperature\\CBT pill}};
\node[process, right=1.4cm of input] (preproc) {
{\footnotesize Outlier removal, resampling,\\time synchronization, normalization\\Light log-transform}};
\node[process, right=1.4cm of preproc] (features) {
{\footnotesize Window statistics: mean, std, slope, min, max, last value}};
\node[process, below=1.2cm of features] (cbt) {Ground Truth Construction\\[2pt]
{\footnotesize Cosinor fit of CBT:\\$T(t)=M+A\cos(\omega t+\phi)$\\$\rightarrow$ Phase $\theta_{CBT}(t)$}};
\node[process, right=1.4cm of features] (model) {Machine Learning Model\\[2pt]
{\footnotesize Random Forest / Gradient Boosting /\\LSTM / CNN-LSTM\\Trained on historical windows}};
\node[process, right=1.4cm of model] (output) {Estimated Circadian Phase\\[2pt]
{\footnotesize $\hat{\theta}(t)$ (0–2$\pi$)\\cross-validation}};

\draw[arrow] (input) -- (preproc);
\draw[arrow] (preproc) -- (features);
\draw[arrow] (features) -- (model);
\draw[arrow] (model) -- (output);
\draw[arrow] (cbt) -- (model);

\node[above=0.15cm of input, font=\bfseries] {Sensor Data Acquisition};
\node[above=0.15cm of preproc, font=\bfseries] {Signal Processing};
\node[above=0.15cm of features, font=\bfseries] {Feature Engineering};
\node[above=0.15cm of model, font=\bfseries] {Model Training};
\node[above=0.15cm of output, font=\bfseries] {Estimation \& Evaluation};

\end{tikzpicture}}
\vspace{1mm}

\caption{Overall methodological workflow for low latency circadian phase estimation from wearable data. 
Signals from wearable and ingestible sensors are preprocessed, synchronized, and transformed into window-based statistical features and sequential inputs. 
The reference phase is derived via cosinor fitting of the CBT. 
Different models are trained on historical input windows to estimate the instantaneous circadian phase $\hat{\theta}(t)$.}

\label{fig:methods_overview}
\end{figure*}

The methodological workflow is illustrated in Fig.~\ref{fig:methods_overview}. Multimodal wearable signals were preprocessed, synchronized, and transformed into window-based features. The reference phase was derived via cosinor fitting of CBT. Machine-learning and deep-learning models were then trained to estimate instantaneous circadian phase using only historical observations.

A central design principle was minimum latency operation: at each time $t$, the model estimated $\hat{\theta}(t)$ using only data within $[t-W,\,t]$, without access to future observations. The term ``real-time'' here refers to this causal constraint rather than the execution platform. The evaluation was conducted retrospectively on pre-collected data using a sliding-window protocol that simulates online deployment. The computational cost of tree-based models is sufficiently low for edge-device inference, but on-device deployment remains a direction for future validation.

\subsection{Signal Preprocessing and Normalization}
\label{sec:preprocessing}
All sensor data streams were preprocessed prior to feature extraction. Extreme outliers were removed using an interquartile range (IQR) criterion to exclude physiologically implausible values. Signals were then resampled to a common one minute resolution and synchronized using a unified UTC time reference to ensure temporal alignment across modalities.

Short missing segments lasting less than five minutes were linearly interpolated to preserve temporal continuity, whereas longer gaps were excluded from downstream analysis.

To reduce inter-individual variability and enable cross-subject generalization, all signals were normalized on a per-participant basis. Specifically, physiological and behavioral signals were z-scored individually for each participant using that participant's own recording statistics. Environmental light intensity was first log-transformed to compress its heavy-tailed distribution and then z-scored per participant. Because the cross-validation scheme partitions data at the participant level (i.e., all data from a given participant appear exclusively in either the training or test set), per-participant normalization does not introduce information leakage between folds.

\subsection{Feature Engineering and Historical Window}

Feature extraction was performed on preprocessed wearable signals. The extracted modalities included environmental light exposure, physical activity (wrist-based motion counts and a net acceleration magnitude $\text{acc}_{\mathrm{net}} = \max(\sqrt{a_x^2+a_y^2+a_z^2}-1,\,0)$ derived from tri-axial accelerometry), heart rate, and skin temperature.

All features were organized using sliding historical windows. At each time $t_i$, the circadian phase $\hat{\theta}(t_i)$ was inferred from observations within $[t_i - W, t_i]$, with $W \in \{30, 60, 120, 240, 480, 1440\}$ minutes. For machine-learning models, each window was summarized using six statistical descriptors: mean, standard deviation, minimum, maximum, last observed value, and linear trend (slope). For sequence-based models, minute-level signals were provided directly as multivariate time-series inputs.

\subsection{Ground Truth}
The CBT served as the reference oscillator for the estimation of the circadian phase.
For each participant and measurement segment, a single-component cosinor model was fitted to the CBT time series to capture the dominant 24-hour rhythmic component as described in \cite{huang2021predicting}.
The phase of the fitted cosinor was treated as the reference circadian trajectory:
\[
T(t)=M+A\cos(\omega t+\phi), \quad \omega = \frac{2\pi}{\SI{24}{h}}.
\]
The fitted acrophase parameter $\phi$ was used to reconstruct the ground-truth phase trajectory:
\[
\theta_{\mathrm{CBT}}(t)=(\omega t+\phi)\bmod 2\pi.
\]
This trajectory represents the instantaneous circadian phase (0--2$\pi$ corresponding to 0--24 h) aligned with each timestamp.

\subsection{Circular Representation of Circadian Phase}

To avoid discontinuities at the 0/24~h boundary, the reference phase $\theta(t) \in [0,2\pi)$ was encoded as $y_{\sin}(t) = \sin(\theta(t))$ and $y_{\cos}(t) = \cos(\theta(t))$. All models were trained to jointly estimate $(\hat{y}_{\sin}, \hat{y}_{\cos})$, with the phase recovered via $\hat{\theta}(t) = \operatorname{atan2}(\hat{y}_{\sin}, \hat{y}_{\cos}) \bmod 2\pi$. For interpretability, phase estimates were converted to hours as $\hat{t} = \frac{24}{2\pi}\hat{\theta}$.

\subsection{Model Architectures and Training Configuration}

Several model classes were evaluated for circadian phase estimation, covering both feature based and sequence based approaches. Random Forest (RF) and Gradient Boosting Regressor (GBR) were adopted as machine learning models that estimate circadian phase from aggregated multimodal features. In addition, sequence based neural network models were considered to capture temporal dependencies within historical windows, including single layer LSTM,and a hybrid CNN-LSTM architecture.

All models were trained and evaluated under identical window lengths and sensor modality configurations to ensure a controlled comparison. The architectural design and optimized parameters of each model are summarized in Table~\ref{tab:model_config}. For tree-based models, hyperparameters were selected via grid search within each cross-validation fold. For neural network models, parameters were optimized using the Adam optimizer with a learning rate of $10^{-3}$ and a mean squared error loss on the sine--cosine targets. A dropout rate of 0.2 was applied after the recurrent layer to mitigate overfitting. Early stopping with a patience of 20 epochs was used based on validation loss monitored on a held-out portion of the training set. All neural network experiments were implemented in PyTorch.

\begin{table}[!htbp]
\centering
\renewcommand{\arraystretch}{1.25}
\caption{Summary of model configurations and training strategies.}
\label{tab:model_config}
\begin{tabular}{p{1.5cm} p{4.1cm} p{1.8cm}}
\hline
\textbf{Model} & \textbf{Key settings} & \textbf{Training} \\
\hline
RF 
& Number of trees: 200, 400; maximum depth: 10 or none 
& Grid search \\
\hline
GBR
& Number of estimators: 200, 400; maximum depth: 2, 3
& Grid search \\
\hline
LSTM 
& One LSTM layer; hidden size: 128, 256; batch size: 128; epochs: 200
& $\mathrm{lr}=10^{-3}$ \\
\hline
CNN-LSTM 
& 1-dimensional convolution (32 channels, kernel size 5) followed by one LSTM layer with 128, 256 units; batch size: 128; epochs: 200
& $\mathrm{lr}=10^{-3}$ \\
\hline
\end{tabular}
\end{table}

\subsection{Modality grouping for ablation analysis}

To evaluate the contribution of different wearable signals, features were grouped into physiologically meaningful modality classes and organized into a cumulative hierarchy.
This hierarchy reflects increasing sensing complexity and data availability in real-world deployments, with the aim of estimating circadian phase using the smallest feasible set of easily obtainable signals.
Based on this design, a structured set of modality configurations was defined for ablation and multimodal analysis, as summarized in Table~\ref{tab:modality_features}.

\begin{table}[!htbp]
\centering
\caption{Overview of modality configurations and corresponding feature compositions.}
\label{tab:modality_features}
\renewcommand{\arraystretch}{1.15}
\begin{tabular}{p{1cm} p{2.5cm} p{4.0cm}}
\hline
\textbf{Modality} & \textbf{Composition} & \textbf{Feature List} \\
\hline
M1 & Light &
\textit{light(lux)} \\
M2 & Activity &
\textit{motion\_counts, acc\_net} \\
M3 & Skin Temperature &
\textit{skin\_temp} \\
M4 & Cardiovascular &
\textit{heart rate} \\
M5 & Light + Activity &
\textit{light(lux), motion\_counts, acc\_net} \\
M6 & M5 + Temperature &
\textit{light(lux), motion\_counts, acc\_net, skin\_temp} \\
M7 & All modalities &
\textit{light(lux), motion\_counts, acc\_net, skin\_temp, heart rate} \\
\hline
\end{tabular}
\end{table}

\subsection{Performance Metrics}
\label{sec:metrics}

Because circadian phase is a circular variable, model performance was evaluated using circular error metrics. Specifically, we define the $q$-th order circular absolute moment of the phase estimation error as
\begin{equation}
\xi_q =
\frac{1}{N}\sum_{i=1}^{N}
\left|
\operatorname{mod}\!\left(\hat{\theta}_{i} - \theta_{i} + \pi,\, 2\pi\right) - \pi
\right|^q,
\end{equation}
where $N$ denotes the total number of evaluated samples, $\hat{\theta}_{i}$ is the estimated circadian phase, and $\theta_{i}$ is the corresponding ground-truth phase for the $i$th sample.

In particular, the circular mean absolute error (CMAE) corresponds to the first-order moment $\xi_1$ and is used as the primary performance metric throughout this study.

To ensure robust performance evaluation and reduce the impact of inter-subject variability, all models were trained and evaluated using 5-fold cross-validation. Model evaluation was restricted to time periods for which ground-truth CBT measurements were available. For each fold, performance metrics were computed on the held-out test set. Reported CMAE values therefore correspond to the mean and standard deviation across the five folds.

\section{Results}
\label{sec:results}

\begin{table*}[t]
\centering
\caption{Circadian phase estimation performance (CMAE) across different sensor modality configurations using a 480-minute historical window. Results are reported for multiple model classes under participant-based 5-fold cross-validation.}
\label{tab:model_performance}
\begin{tabular}{lcccc}
\toprule
\textbf{Modality} & \textbf{RF} & \textbf{GBR} & \textbf{LSTM} & \textbf{CNN--LSTM} \\
\midrule
M1 (Light) 
& $1.40 \pm 0.25$ & $1.26 \pm 0.21$ & $1.65 \pm 0.48$ & $1.36 \pm 0.20$ \\
M2 (Activity) 
& $1.79 \pm 0.40$ & $1.81 \pm 0.33$ & $2.66 \pm 0.41$ & $2.73 \pm 0.37$ \\
M3 (Skin Temperature) 
& $5.54 \pm 0.75$ & $5.55 \pm 0.96$ & $5.68 \pm 0.71$ & $5.58 \pm 0.70$ \\
M4 (Cardiovascular) 
& $3.56 \pm 1.48$ & $3.33 \pm 1.15$ & $4.33 \pm 0.78$ & $4.32 \pm 0.83$ \\
M5 (Light + Activity) 
& \textbf{1.19 $\pm$ 0.18} & $1.19 \pm 0.22$ & $2.31 \pm 0.76$ & $2.22 \pm 0.76$ \\
M6 (Light + Activity + Temp) 
& $1.24 \pm 0.23$ & $1.25 \pm 0.26$ & $2.27 \pm 0.58$ & $2.81 \pm 0.26$ \\
M7 (All modalities) 
& $1.21 \pm 0.25$ & $1.28 \pm 0.27$ & $2.32 \pm 0.53$ & $2.64 \pm 0.30$ \\
\bottomrule
\end{tabular}
\end{table*}

\subsection{Overall performance} 

Table~\ref{tab:model_performance} summarizes circadian phase estimation performance across different sensor modality sets using a 480-minute window under participant-based 5-fold cross-validation. 

Under single modality settings, light consistently yielded the best performance, followed by activity, while skin temperature and cardiovascular signals resulted in substantially higher errors. For tree-based models, combining light and activity (M5) led to the best overall performance, achieving a CMAE of approximately 1.19~h, with standard deviations of 0.18 and 0.22 under 5-fold cross-validation for RF and GBR, respectively.

In contrast, sequence-based neural networks did not exhibit systematic performance gains with increasing modality combinations, and in some cases performed worse than the single-light configuration. Notably, the CNN--LSTM model already achieved strong performance under the light-only modality, with a CMAE of approximately 1.36~h, better than its multimodal counterparts.

\subsection{Effect of window length}

\begin{figure}[!ht]
    \centering
    \includegraphics[width=\linewidth]{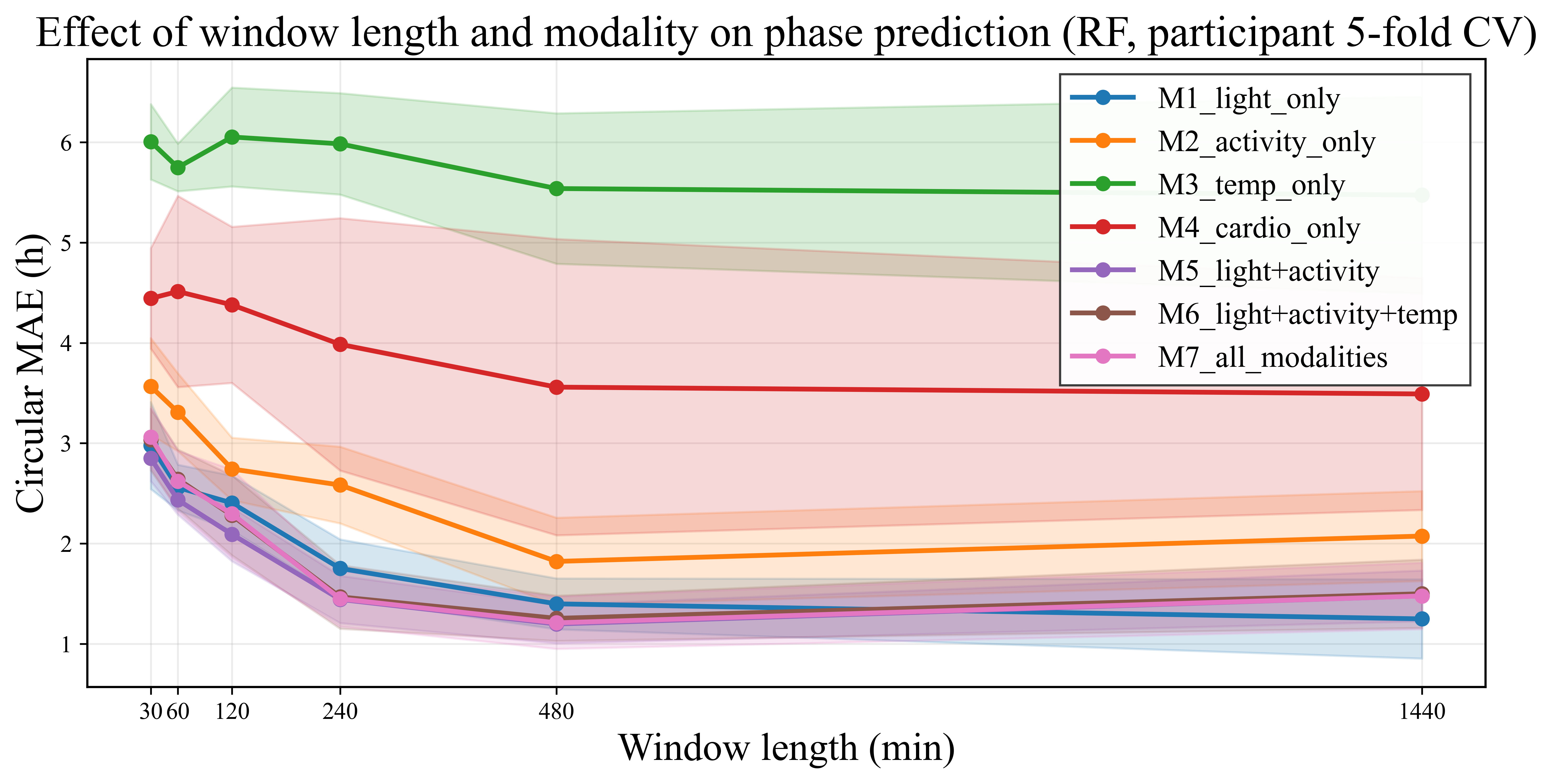}
    \caption{
        Effect of window length and sensor modality on circadian phase estimation accuracy 
        using Random Forest models with 5-fold cross-validation. 
    }
    \label{fig:window_length_mae}
\end{figure}

The effect of historical window length was examined using the RF model across modality configurations (Fig.~\ref{fig:window_length_mae}). Estimation accuracy improved monotonically from 30 to 480 minutes, after which performance saturated for nearly all modality combinations. Beyond 480 minutes, further enlarging the window yielded no significant benefit.

To further quantify this effect, we analyzed the distribution of estimation errors across all evaluated samples, as summarized in Table~\ref{tab:M5_window_performance}. Using a 480-minute window, 53.97\% of phase estimates fall within 1 hour of the reference phase, and 83.10\% fall within 2 hours, representing the best overall performance among all tested window lengths.

\begin{table}[!ht]
\centering
\caption{Performance of Random Forest with M5 (Light + Activity) under different window lengths.}
\begin{tabular}{cccc}
\toprule
\textbf{Window [min]} & \textbf{CMAE [h]} & \textbf{Within-1h [\%]} & \textbf{Within-2h [\%]} \\
\midrule
30   & 2.90 & 23.62      & 46.93     \\
60   & 2.42 & 28.87      & 53.93    \\
120  & 2.08 & 39.48      & 61.98     \\
240  & 1.46 & 49.40      & 75.27     \\
480  & \textbf{1.19} & \textbf{53.97}     & \textbf{83.10}     \\
1440 & 1.48 & 48.23      & 72.23     \\
\bottomrule
\end{tabular}
\label{tab:M5_window_performance}
\end{table}

\subsection{Model comparison}

\begin{figure}[!ht]
    \centering
    \includegraphics[width=\linewidth]{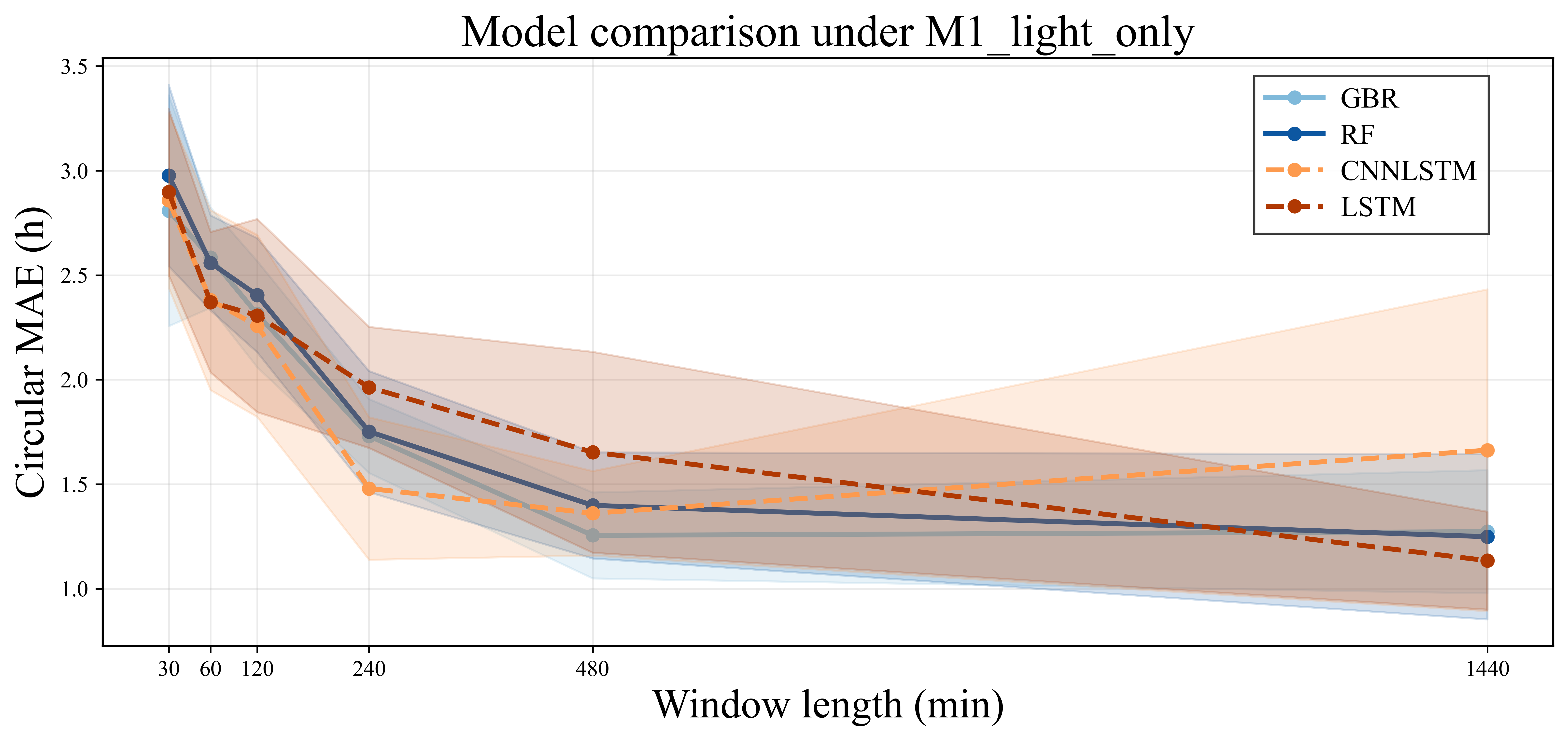}
    \caption{Model comparison under the M1 (light-only) modality across different window lengths. Circular mean absolute error (CMAE) is reported using participant-level 5-fold cross-validation. Different models are evaluated over window lengths ranging from 30 to 1440 minutes, with shaded regions indicating standard deviation across folds.}
    \label{fig:M1_model_comparison}
\end{figure}

Figure~\ref{fig:M1_model_comparison} compares different model classes under the M1 (light-only) configuration across varying window lengths. Overall, all models benefit from increasing historical context, with accuracy improving as window length increases.

For tree-based models (RF and GBR), performance improves steadily up to approximately 480 minutes, beyond which accuracy saturates. In contrast, sequence-based models exhibit different trends: the LSTM model shows a more gradual and sustained improvement with longer windows, whereas the CNN--LSTM achieves strong performance at shorter windows, with particularly competitive accuracy around 240 minutes, indicating effective use of local temporal patterns.

\subsection{Case Study}

\begin{figure}[!htbp]
    \centering
    \includegraphics[width=\linewidth]{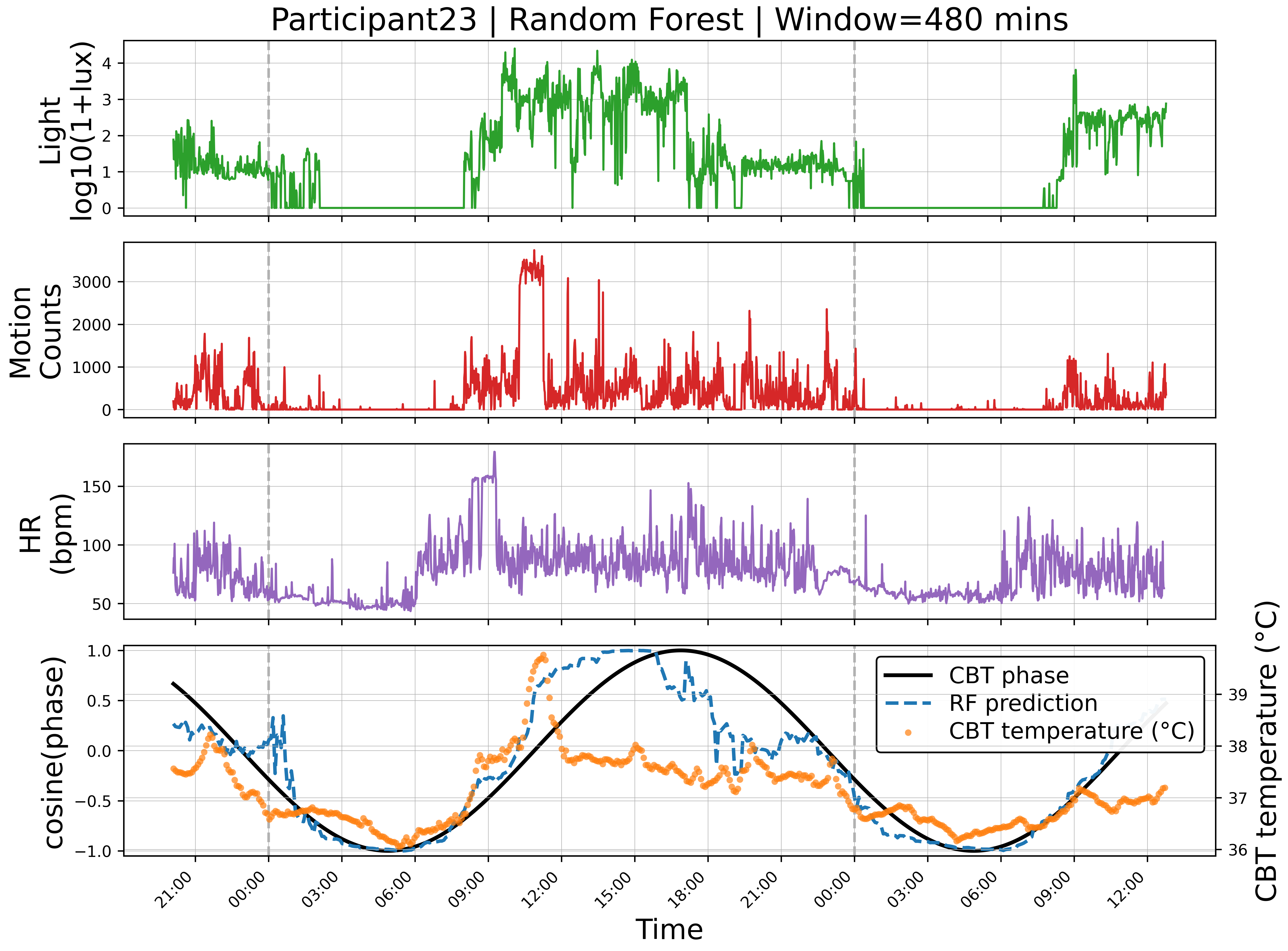}
    \caption{
    Example qualitative results for Participant\#23 using the RF model with M5 (light + activity) features.
    The top three panels show the wearable signals, including light intensity, motion counts, and heart rate.
    The bottom panel compares the reference circadian phase (black solid line) and the RF estimated phase (blue dashed line), overlaid with the corresponding CBT measurements (orange dots, right axis).}
    \label{fig:qualitative_case_participant23}
\end{figure}

\begin{figure}[!htbp]
    \centering
    \includegraphics[width=\linewidth]{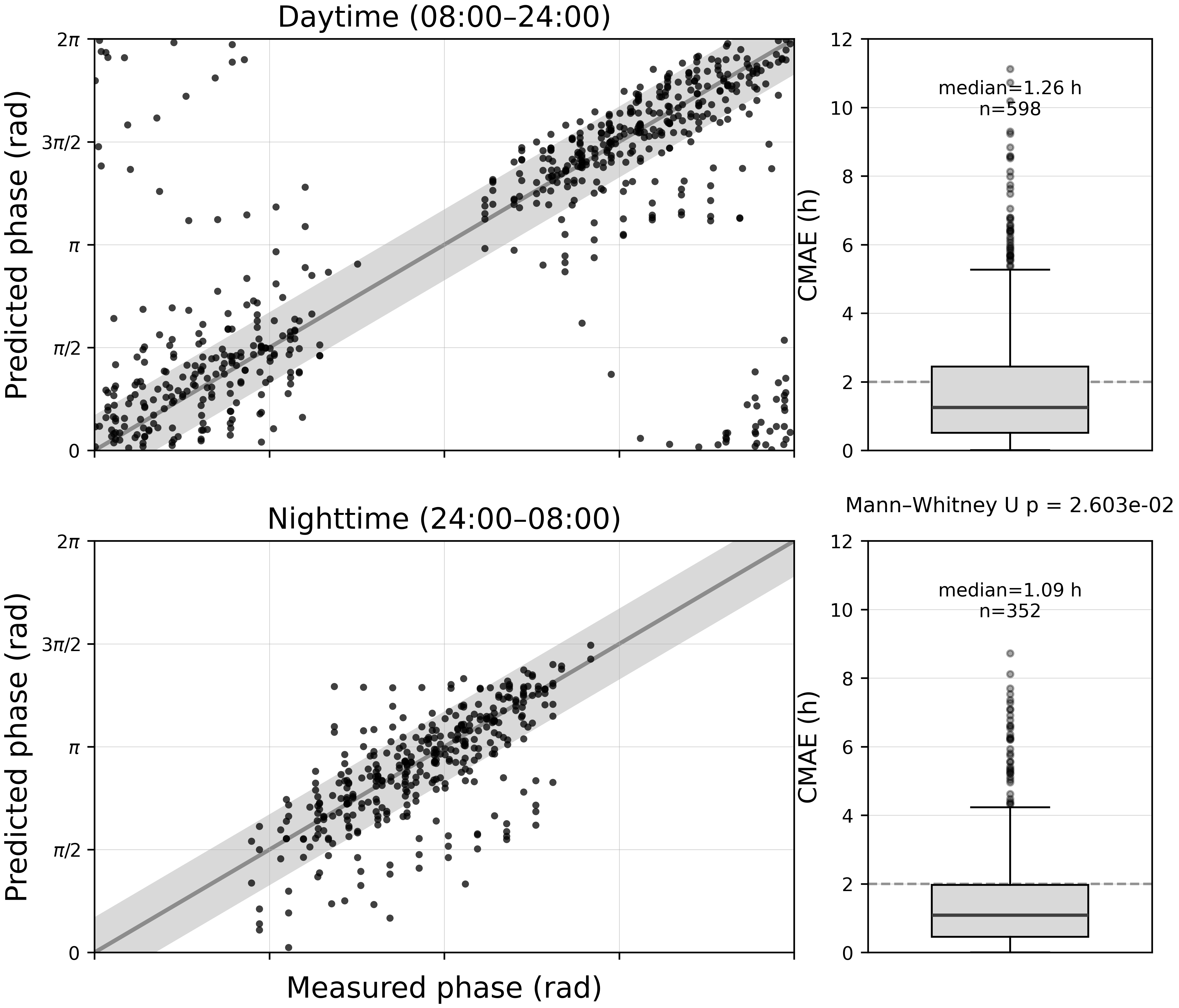}
    \caption{Measured versus estimated circadian phase under daytime and nighttime conditions using the RF model with M5 (light + activity) features and a 480-minute window. Results are aggregated across all participants. The solid diagonal indicates the identity line, and the shaded region denotes a 2-hour estimation error.}
    \label{fig:scatter_day_and_night}
\end{figure}

Fig.~\ref{fig:qualitative_case_participant23} shows a representative case study for Participant~23 using leave-one-out (LOO) training. The estimated phase closely follows the CBT rhythm, with more stable estimates during nighttime and larger deviations during daytime periods of elevated activity. To assess whether this pattern generalizes, the LOO protocol was applied across all participants, with errors analyzed separately for daytime (08:00--24:00) and nighttime (24:00--08:00). As shown in Fig.~\ref{fig:scatter_day_and_night}, nighttime errors are significantly lower (Mann--Whitney U test, $p = 0.026$), confirming that estimation is more reliable during periods of reduced behavioral variability.

\section{Discussion}

The results should be interpreted in the context of free-living data collection, where wearable signals are subject to noise, missing segments, and outliers. This is particularly pronounced for skin temperature and heart rate, which are sensitive to motion, posture, and sensor artifacts. The poor performance of skin temperature (CMAE $>$ 5.5~h), despite its known circadian modulation~\cite{Reid2019}, can be attributed to the distal sensor's susceptibility to environmental conditions, vasomotor artifacts, and contact issues during ambulatory wear. While skin temperature gradients are a well-established circadian marker under controlled laboratory conditions, behavioral masking in free-living settings substantially degrades its circadian signal-to-noise ratio.

For each window, features were summarized as single statistical values, which improves robustness but discards temporal structure. Alternative strategies---such as multi-resolution statistics or frequency-domain descriptors---could better exploit temporal information and merit future investigation. The loss of fine-grained temporal structure may explain the observed performance saturation at intermediate window lengths. Sequence-based models, while suited for temporal dependencies, may struggle with long noisy sequences under limited training data and simple architectures. The degradation of deep learning models with additional modalities likely reflects the difficulty of learning cross-modal representations from only 14 participants, where increased dimensionality exacerbates overfitting. Tree-based models are inherently more robust to noisy features due to built-in feature selection.

The single-component cosinor fit of CBT was adopted as the ground-truth reference because the CBT nadir is closely linked to the endogenous pacemaker~\cite{Klerman2002,Reid2019}, and the cosinor model acts as a smoothing filter attenuating behavioral noise. Nevertheless, CBT in free-living settings is affected by masking from activity, meals, and posture~\cite{Byrne2007}, which can locally distort the fitted rhythm and introduce reference uncertainty. Part of the observed error may therefore reflect reference limitations. Future work could benefit from multi-marker strategies or masking purification techniques.

Although our results indicate that approximately 8 hours of historical data are sufficient to achieve good estimation performance, it should be noted that all input signals were normalized on a per-subject basis. In practical deployment scenarios, this may require additional data to obtain reliable normalization statistics.

The sample of 14 participants is comparable to prior free-living circadian studies~\cite{Stone2019,Suarez2021CircadianPhasePrediction,Weed2025ParticleFiltering}, and participant-based cross-validation provides a conservative generalization estimate. Nonetheless, validation on larger and more diverse cohorts remains an important future direction.

The advantage of tree-based models over LSTM and CNN--LSTM should be interpreted with caution. Deep learning models require more training data, and the simple architectures used here may not fully exploit sequence modeling capacity. We do not claim tree-based models are inherently superior, but rather that they offer a more data-efficient solution in the small-sample regime of current free-living studies. This study also did not benchmark against established circadian models such as limit-cycle oscillators~\cite{Hannay2020} or particle filters~\cite{Bonarius2021ParticleFilterCircadian,Weed2025ParticleFiltering}, which typically require longer observation windows, making direct comparison under the short-window constraint difficult.

Light exposure alone (M1) performs competitively with multimodal combinations, consistent with light being the dominant Zeitgeber~\cite{Czeisler1989}. Adding activity (M5) yielded a modest improvement from 1.26~h (GBR) to 1.19~h (RF), while further additions (M6, M7) produced no consistent gains. Although the M1--M5 difference is small, it was consistent across models and folds. A formal statistical comparison was not conducted due to the limited number of folds ($k=5$). For most chronotherapy and healthy-building applications operating on multi-hour timescales, differences below 0.5~h are unlikely to be clinically meaningful.

Taken together, these observations highlight key challenges for circadian phase estimation in free-living settings, where data quality limitations, sample size constraints, and system-level requirements interact. In such settings, the use of long historical recordings or highly complex models may be impractical, motivating approaches that explicitly consider latency, robustness, and resource limitations during model design.

\section{Conclusion}

This paper presented a low latency framework for circadian phase estimation from wearable data under free-living conditions. Using light exposure and physical activity, the approach achieved a CMAE of 1.19~h with approximately 8 hours of historical data, beyond which accuracy saturates. Compared with methods requiring full circadian cycles, this substantially reduces estimation latency and computational requirements, making it suitable for edge-based deployment.

Future work will focus on improving robustness to free-living noise, validating on larger cohorts, incorporating richer temporal feature representations, and extending the framework toward adaptive circadian intervention systems.

\section*{Acknowledgements}
This work was supported by the Project LoLiPoP-IoT (Long Life Power Platforms for Internet of Things), grant agreement No.~101112286. The authors thank Mauritius van Maurik for his contributions to data cleaning and preprocessing, and Jan Bergmans for providing constructive comments and valuable suggestions.

\bibliographystyle{IEEEtran}
\bibliography{references}

\end{document}